\documentclass[pra,showpacs,preprintnumbers,amsmath,amssymb,tightenlines,twocolumn,superscriptaddress]{revtex4}

\usepackage{graphicx}
\usepackage{dcolumn}
\usepackage{bm}
\usepackage{epsfig}
\usepackage{appendix}
\usepackage{stmaryrd}
\usepackage{placeins}

\newcommand{\ket}[1]{|\,{#1}\,\rangle}
\newcommand{\braket}[2]{\mbox{$\langle\,{#1}\, | \,{#2}\,\rangle$}}

\newcommand{\expec}[1]{\langle #1 \rangle}
\newcommand{\mods}[1]{\left| #1 \right| ^{2}}

\newcommand{\s}{\vspace{0.2cm}}
\newcommand{\xm}{\vspace{0.35cm}}

\newcommand{\sub}[2]{{#1}_{\mbox{\!\! \scriptsize #2}}}

\def\dl{\tilde}
\def\half{\frac{1}{2}}

\def\beq{\begin{equation}}
\def\eeq{\end{equation}}

\newcommand{\rref}[1]{ref.~\cite{#1}}
\newcommand{\fref}[1]{Fig.~\ref{#1}}
\newcommand{\frefp}[2]{Fig.~\ref{#1}~(#2)}

\newcommand{\eref}[1]{Eq.~(\ref{#1})}
\newcommand{\esref}[2]{Eqs.~(\ref{#1}) and (\ref{#2})}

\newcommand{\sref}[1]{section~\ref{#1}}

\newcommand{\cref}[1]{chapter~\ref{#1}}
\newcommand{\Sref}[1]{Section~\ref{#1}}
\newcommand{\Cref}[1]{Chapter~\ref{#1}}

\newcommand{\aref}[1]{appendix~\ref{#1}}
\newcommand{\bref}[1]{(\ref{#1})}

\usepackage{hyperref}
\usepackage{bbm}
\usepackage{MnSymbol}
\usepackage{afterpage}
\makeatother

\global\long\def\phbraces#1{\mathinner{[{#1}]}}
\global\long\def\pkeys#1{\mathinner{\{{#1}\}}}
\global\long\def\pbars#1{\mathinner{|#1|}}

\global\long\def\ket#1{\mathinner{\lvert#1\rangle}}
\global\long\def\braket#1#2{\mathinner{\langle{#1\lvert#2}\rangle}}

\global\long\def\heff{\sub{\hbar}{eff}}
\global\long\def\g4{g^{(4)}}
\global\long\def\ad{\hat{a}^{\dagger}}

\global\long\def\a{\hat{a}}

\begin{document}

\title{Dynamical tunnelling of a Nano-mechanical Oscillator}
\author{Piyush Jangid}
\affiliation{Department of Physics, Indian Institute of Science Education and Research, Bhopal, Madhya Pradesh 462 066, India}
\author{Anil~{Kumar Chauhan}}
\affiliation{Department of Physics, Indian Institute of Science Education and Research, Bhopal, Madhya Pradesh 462 066, India}
\affiliation {Department of Optics, Palack\'y University, 17.~listopadu 1192/12, 77146 Olomouc, Czechia}
\author{Sebastian~W\"uster}
\affiliation{Department of Physics, Indian Institute of Science Education and Research, Bhopal, Madhya Pradesh 462 066, India}
\email{sebastian@iiserb.ac.in}
\begin{abstract}
The study of the quantum to classical transition is of fundamental as well as technological importance, and focusses on mesoscopic devices, with a size for which either classical physics or quantum physics can be brought to dominate. A particularly diverse selection of such devices is available in cavity quantum-optomechanics. We show that these can be leveraged for the study of dynamical-tunnelling in a quantum chaotic system. This effect probes the quantum to classical transition deeply, since tunnelling rates sensitively depend on the ability of the quantum system to resolve the underlying classical phase space. We show that the effective Planck's constant, which determines this phase space resolution, can be varied over orders of magnitude as a function of tunable parameters in an opto-mechanical experiment. Specifically, we consider a membrane-in-the-middle configuration of a mechanical oscillator within an optical cavity, where the intracavity field is modulated periodically by the external laser source. We demonstrate that a mixed regular and chaotic phase space can be engineered in one spatial dimension, through a significant quartic opto-mechanical interaction. 
For that case, we explore the expected dynamical tunnelling rates using Floquet theory and map out values of the effective Planck's constant that should be within practical reach.
\end{abstract}

\maketitle

\section{Introduction}
%
Through achievements such as the cooling of ever more macroscopic oscillators to the quantum-mechanical ground-state \cite{Cornell2010:gsmotiondetection, Caltech2011:nanomechosccooling, MIT2011:atomicensamblecooling, EPFL2011:qmgscooling, Boulderl2011:sidebandqmgscooling, delic2020cooling}, quantum opto-mechanics \cite{kippenberg:review,poot:mechquantumsyst,aspelmeyer:review} has established itself as a leading discipline for the exploration of the quantum to classical transition \cite{Schlosshauer_decoherence_review,buchmann:mediated_int}. At the heart of this progress is the intricate control over light-matter interaction, which also facilitates quantum information transfer between different spectral realms \cite{Barzanjeh_2011_optical,Barzanjeh_2012_optical,Bochmann_Cleland:microwave_optical,Andrews_2014,Bagci_2014_detection}, the generation of non-classical states of light \cite{Palomaki_osc_mw_entanglement_science,Riedinger_nonclass_mechlight_nature} and oscillators \cite{clerk2008back,mari2009gently,wollman2015quantum}, interfacing of light, mechanics and cold atoms in hybrid systems \cite{singh:wignerntomog,singh:cantilevermol,sanzmora:mirror_classical,chauhan2016atom,chauhan2017motion,hoikwan:mirror_quantum} or state tomography \cite{vanner2011pulsed,yin2013large,lecocq2015quantum,vanner2015towards,lei2016quantum,sanzmora_QNDtomography}.

\begin{figure}[htb]
\centering 
\includegraphics[width=\linewidth]{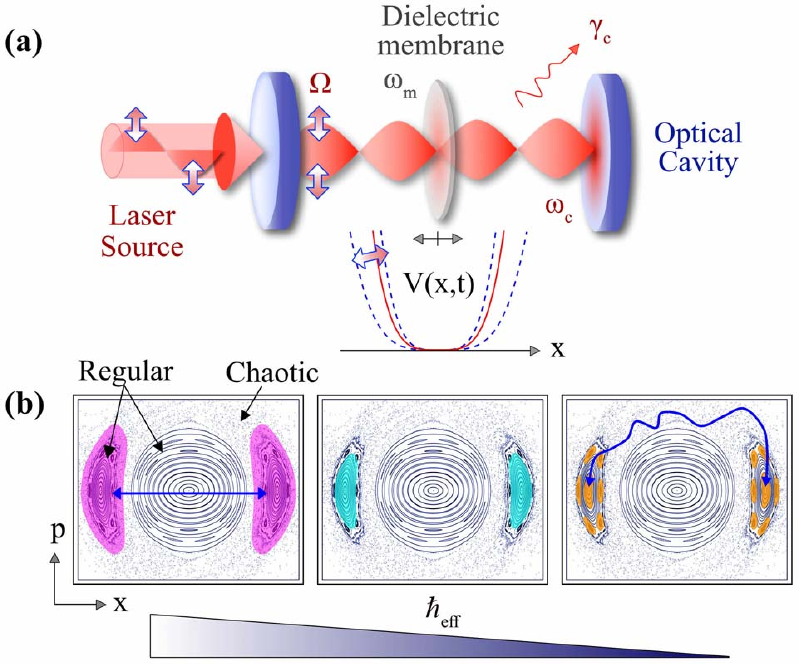}
\caption{(a) Membrane-in-the-middle within a cavity as quantum harmonic oscillator with tunable, light-driven anharmonicity of the potential $V(x,t)$. The typical modulation range of the oscillator potential is sketched below. A fairly large cavity decay-rate $\gamma_c$ makes sure cavity dynamics follows the external drive modulation. (b) Stroboscopic Poincar{\'e} section of this system with contours of the Husimi function of a Floquet state (magenta, cyan, orange) involved in dynamical tunnelling for decreasing effective Planck's constant $\sub{h}{eff}$ as shown.
\label{system_sketch}}
\end{figure}
Many of these applications and others envisaged for the future hinge on a non-linear coupling of the mechanical motion to the  light \cite{khosla2013quantum,vanner2013cooling,doolin2014nonlinear,brawley2016nonlinear,bennett2016quantum, albarelli2016nonlinearity,leijssen2017nonlinear}. Designing devices with ever larger non-linear coupling strengths, is hence an intensively pursued activity in the field \cite{weiss2019quantum,rakhubovsky2019stroboscopic,delic2020levitated,cattiaux2020beyond}. Here, we demonstrate that a strong quartic opto-mechanical interaction also benefits engineering light-controlled non-linear potentials for a one-dimensional harmonic oscillator, which then becomes a useful platform to explore quantum chaos \cite{book:reichl}. Frequently, problems in quantum chaos involve a mixed phase space containing regular as well as chaotic regions. A phase-space with one spatial dimension can only exhibit chaos if the potential is an-harmonic and the Hamiltonian time-dependent. We will show that an-harmonicity can be provided by the quartic opto-mechanical interaction and time-dependence by a modulation of the light field. Pushing the system across the quantum-classical transition then requires the effective Planck's constant $\sub{h}{eff}$ that arises as commutator between position and momentum operator in some suitably chosen scaled units, to be widely tunable. Since $\sub{h}{eff}$ controls the size of the smallest structures in phase-space that a quantum system can resolve, as it is lowered, finer and finer details of phase-space may become relevant.

We specifically focus on a mechanical oscillator in a membrane-in-the-middle (MIM) setup, as sketched in \frefp{system_sketch}{a}, where a dielectric membrane is placed inside an optical cavity precisely at the position of a node in the field of the relevant cavity mode. It has been experimentally demonstrated in \rref{sankey:mim}, that through careful alignment of the membrane and use of the transverse field structure of cavity modes, a configuration can be found where the usual quadratic coupling between the mechanical oscillator and the light vanishes, and hence the quartic term becomes the leading order of the relevant Taylor expansion. We further assume a fairly lossy cavity, so that its light content can quickly adjust to the power of the drive laser \cite{brawley2016nonlinear,bennett2016quantum,leijssen2017nonlinear}, and thus can be periodically modulated in time. The scheme thus provides a light controlled quartic potential for the membrane, based on radiation pressure. 

As target problem in quantum chaos, we focus in this article on the phenomenon of  dynamical tunnelling. While conventional quantum tunnelling refers to dynamics that is forbidden in classical physics for energetic reasons, dynamical tunnelling refers to dynamics forbidden by symmetry. The phenomenon was first discovered in molecular physics \cite{heller:dt} and has since been investigated also with light in optical cavities \cite{exp:jens,exp:shinohara,exp:ketzmerick}, cold atoms \cite{exp:hensinger,exp:raizen}, microwave resonators \cite{exp:dembowski,exp:backer} or electrons in quantum dots \cite{exp:bird}. Dynamical tunnelling rates sensitively depend on the degree to which the quantum system can resolve the classical phase space \cite{utermann:dtperiod,narimanov:heffrole}, which manifests itself for example through changes of these rates by orders of magnitude as the system becomes sensitive to the presence of higher-order resonance island chains \cite{luca:nonlinresonance,exp:ketzmerick}. To explore such features on a single experimental platform, being able to tune the importance of quantum effects via some effective Planck's constant $\sub{h}{eff}$ through appropriate choice of scales in the system is essential.

For the opto-mechanical setup discussed above, we suggest a suitable choice of these scales and explore in detail how widely $\sub{h}{eff}$  can then be varied. Prior to that, we explore the variation of the classical phase space for the driven anharmonic oscillator as a function of driving parameters, and demonstrate with a few examples how dynamical tunnelling would be manifest and tunable in such a system. 

This article is organized as follows: \Sref{anh_osc_system} presents our model system, and explores the quantum-classical phase space correspondence for it. \Sref{dyntun} shows exemplary simulations of dynamical tunnelling in the opto-mechanical setup. We then demonstrate how the required initial states for dynamical tunnelling could be practically approximated in \sref{inistate} and then show our main results in \sref{ps_resolution}, where we survey to what extent the effective Planck's constant $\heff$ can be tuned in the proposed setting. Finally, \sref{conc_out} gives a conclusion and outlook. 
%
\section{Nanomechanical oscillator with driven anharmonicity}
\label{anh_osc_system}

We consider a nano-mechanical oscillator (membrane) suspended inside a laser-driven optical cavity, shown in \frefp{system_sketch}{a}. A mechanical mode with frequency $\omega_m$ of the oscillator is coupled only quartically to a cavity mode with frequency $\sub{\omega}{c}$, i.e.~the cavity mode frequency depends quartically on the displacement of the membrane. This is possible only under specific design conditions such as discussed in \rref{sankey:mim}, involving a tilt in the membrane through a few milliradians at a node or antinode of the cavity field.

The Hamiltonian describing this system is
\begin{equation} \label{eq:H_system}
    \sub{\hat{H}}{sys} =  \hat{H}_c+ \hat{H}_m + \sub{\hat{H}}{int}.
\end{equation}
Here the Hamiltonian of the driven cavity field is
\begin{equation} \label{eq:H_cavity}
	  \hat{H}_c= \hbar \omega_c \ad \a + i\hbar \zeta (\ad e^{-i\sub{\omega}{$\ell$} t} - \mbox{H.c.}),
\end{equation}
where $\ad$ ($\a$) is the bosonic creation (annihilation) operator of a cavity mode photon, and the cavity is externally driven by a laser field of frequency $\sub{\omega}{$\ell$}$ and amplitude $\zeta = \sqrt{2 \sub{P}{$\ell$} \gamma_c/\hbar\sub{\omega}{$\ell$}}$. In the latter, $\sub{P}{$\ell$}$ is the laser power and $\sub{\gamma}{c}$ is the cavity decay-rate. The Hamiltonian of the mechanical oscillator of effective mass $m$ is
\begin{equation} \label{eq:H_mech}
	 \hat{H}_m= \frac{\hat{p}^2}{2m} + \frac{1}{2} m\omega_m^2\hat{x}^2, 
\end{equation}
where the position operator $\hat{x}$ and momentum operator $\hat{p}$ satisfy $\phbraces{\hat{x},\hat{p}} = i\hbar$ as usual, and the second term describes the harmonic potential arising from the mechanical support of the membrane. Most importantly, 
\begin{equation} \label{eq:H_int}
    \sub{\hat{H}}{int} = \hbar \g4 \ad \a \:\: \hat{x}^4
\end{equation}
is the optomechanical interaction Hamiltonian, where $\g4 = (1/4!) \hspace{1pt} \partial^4 \sub{\omega}{c}/\partial x^4$ denotes the quartic dispersive optomechanical coupling strength as discussed in \cite{sankey:mim}.

In a frame rotating at the drive laser frequency $\sub{\omega}{$\ell$}$, the total Hamiltonian becomes
\begin{align} \label{eq:H_int_system_rotating}
    \sub{\hat{H}}{sys} = \hbar \delta_c \ad \a + \frac{\hat{p}^2}{2m} +\frac{1}{2} m\omega_m^2\hat{x}^2 &+ \hbar \g4 \ad \a \:\: \hat{x}^4 \nonumber\\
    &+ i\hbar \zeta (\ad - \a),
\end{align}
where $\delta_c = \omega_c - \sub{\omega}{$\ell$}$ is the cavity-laser detuning.

For the simplest case, in which the cavity decay-rate $\gamma_c$ is large compared to all other relevant scales, the cavity field simply adiabatically follows the external drive, see \aref{cavity_class_app}. Expressing the photon operators via $\a(t) = \alpha(t) + \delta \a(t)$ as their mean field $\alpha(t)\in \mathbb{C}$ and fluctuations $\delta \a(t)$ around the mean value, and for now neglecting fluctuations, \bref{eq:H_int_system_rotating} then simply turns into
\begin{equation} \label{eq:H_osc}
     \sub{\hat{H}}{sys} =\frac{\hat{p}^2}{2m} +\frac{1}{2} m\omega_m^2\hat{x}^2+ \hbar \g4 \mods{\alpha(t)} \hat{x}^4, 
\end{equation}
which describes an anharmonic oscillator. Here, the anharmonicity is not intrinsic to the oscillator, but instead caused by its interaction with the optical field. Through this, the quartic part of the potential can be externally modulated. We assume the form
\begin{equation}
\label{eq:modullight}
	\mods{\alpha(t)} = \mods{\alpha_0} + \mods{A} \cos(\Omega t),
\end{equation}
where $\alpha_0$ is the mean cavity field amplitude, see \aref{eq:alpha0_A}, $A$ is the modulation amplitude and $\Omega$ is the modulation frequency. 

Inserting \bref{eq:modullight} into \bref{eq:H_osc}, the resultant overall mechanical potential becomes $V(x,t)= m \omega_m^2 x^2/2 + \hbar \g4 \mods{\alpha_0} x^4 + \hbar \g4 \mods{A} \cos(\Omega t) x^4$, the modulation of which between the extrema is illustrated in \frefp{system_sketch}{a}. 

We now define a time scale $\tau=\Omega^{-1}$ and a length scale $\mathcal{L}=\left(\sigma \sqrt{8\g4\mods{\alpha_0}/\omega_m} \right)^{-1}$ for the problem, which render the time-dependent Schr{\"o}dinger equation that follows from \bref{eq:H_osc} dimensionless.
As shown in \aref{nondim}, the corresponding effective Hamiltonian is then
\begin{equation} \label{eq:H_osc_t_rescale}
    \hat{H} = \frac{\hat{p}^2}{2} + \kappa \frac{\hat{x}^2}{2} + \kappa \big[ 1+\epsilon \cos(t) \big] \frac{\hat{x}^4}{4} ,
\end{equation}
where
\begin{equation} \label{kappa_epsilon}    
	\kappa = \frac{\omega_m^2}{\Omega^2} \hspace{10pt} \text{and} \hspace{10pt} \epsilon = \frac{\mods{A}}{\mods{\alpha_0}}
\end{equation}
are dimensionless parameters that describe the strength of the quadratic plus quartic potential and the strength of its modulation, and $\sigma = \sqrt{\hbar/2m\omega_m}$ is the zero-point fluctuation amplitude of the mechanical oscillator. The driven anharmonic oscillator potential $V(x,t)=\kappa x^2/2 + \kappa[ 1+\epsilon \cos(t) ]x^4/4$ in \bref{eq:H_osc_t_rescale} is sketched in \frefp{system_sketch}{a} for $\epsilon=0.7$.

Now, $\hat{x}$ and $\hat{p}$ are new dimensionless position and momentum operators for the membrane expressed at the new scales, $\mathcal{L}$, $\tau$, satisfying
\begin{equation} \label{eq:commutation_heff}
	\phbraces{\hat{x},\hat{p}}  \equiv i\heff = i\frac{16\sigma^4 \g4 \mods{\alpha_0}}{\Omega},
\end{equation}
where $\heff$ is the effective Planck constant.  Importantly, since \eref{eq:H_osc_t_rescale} constitutes a one-dimensional, an-harmonic, driven Hamiltonian, it remains relatively simple while still being able to exhibit quantum chaotic behaviour.

For cases where $\kappa$ and $\epsilon$ are of order unity, it is clear that the main features in the phase-space for \bref{eq:H_osc_t_rescale} will also arise around $x,p\sim {\cal O}(1)$.
Since \bref{eq:commutation_heff} controls Heisenberg's uncertainty relation at the new scales, $\heff$ will govern the effective coarse-graining of this phase-space imposed by quantum mechanics. By lowering  $\heff$ through a choice of the parameters in \bref{eq:commutation_heff}, more and more resolution can be obtained.
 
Note, that at this point the choice of $\tau$ and $\mathcal{L}$ and hence resultant expressions for $\heff$ are fairly arbitrary, many other choices are possible. Whether a given choice is of practical utility then hinges on whether it is practically feasible to initiate and interrogate quantum dynamics in the resultant interesting parts of phase-space at the scales chosen.

\subsection{Classical Phase Space}
\label{classical_ps}

\begin{figure}[tb]
\centering
\includegraphics[width=\linewidth]{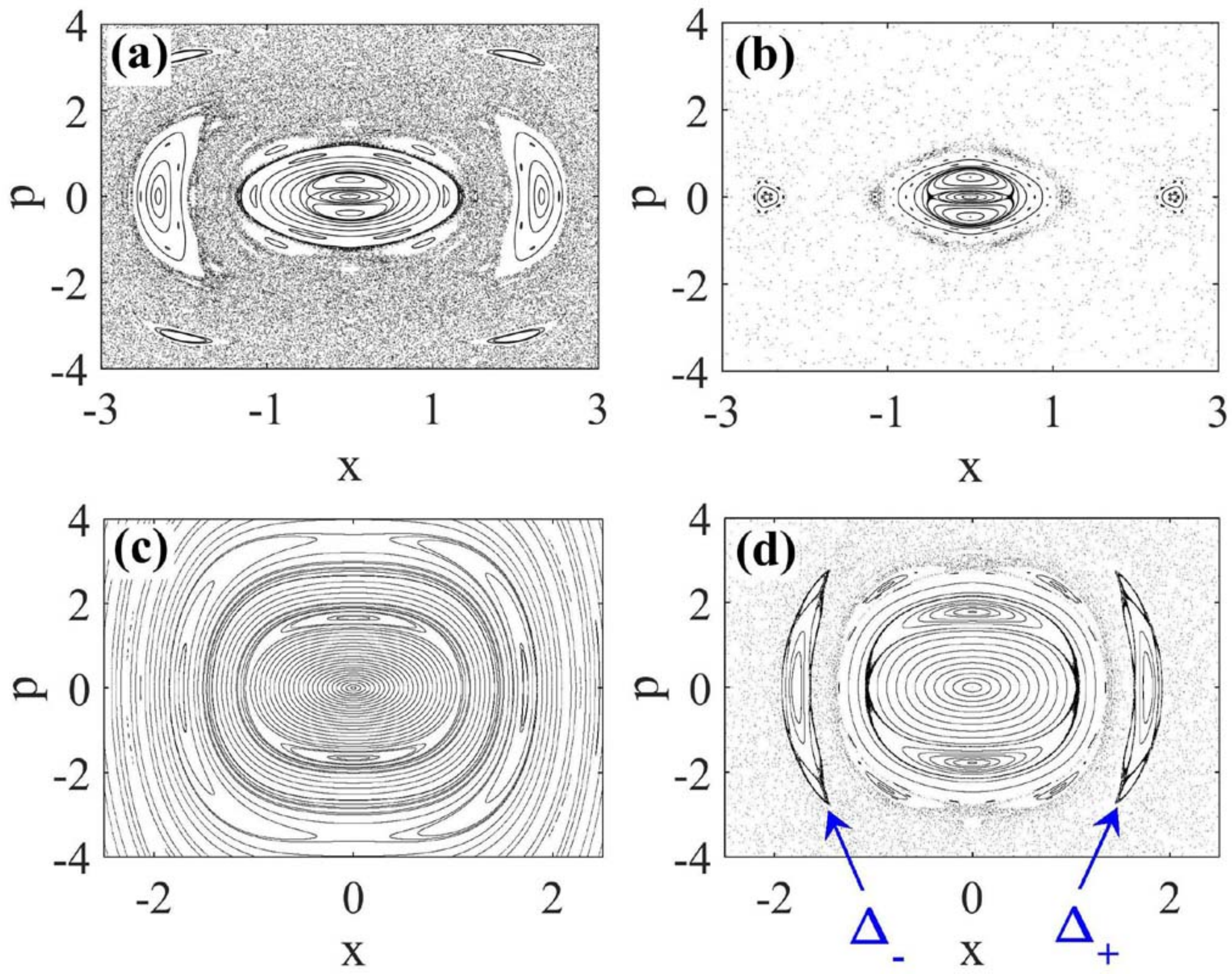}
\caption{Varying phase-space of the classical version of Hamiltonian \bref{eq:H_osc_t_rescale} with the potential strength $\kappa$ and modulation amplitude $\epsilon$ chosen as  (a) $\kappa = 0.2$, $\epsilon = 0.7$, (b) $\kappa = 0.2$, $\epsilon = 0.9$, (c) $\kappa = 1.2$, $\epsilon = 0.7$, and (d) $\kappa = 1.2$, $\epsilon = 0.9$. Labels $\Delta_\pm$ mark period one islands of stability. All panels show a stroboscopic Poincar\'e section as discussed in the text.
\label{phase_space}}
\end{figure}

To demonstrate the utility of the above system for exploration of the quantum to classical transition, let us begin by mapping out the phase space of the Hamiltonian \bref{eq:H_osc_t_rescale} when viewed classically. The classical equations of motion corresponding to this Hamiltonian system are
\begin{align}
	\dot{x} &= p, \label{eq:x_dot}\\
	\dot{p} &= -\kappa x -\kappa[1+\epsilon \cos(t)]x^3. \label{eq:p_dot}
\end{align}
To visualize phase space, we look at its stroboscopic Poincar\'e sections as sketched in \frefp{system_sketch}{b} and shown in \fref{phase_space}. In these figures, solutions of \esref{eq:x_dot}{eq:p_dot} from a large range of initial conditions of $x$ and $p$ are plotted stroboscopically, i.e., at times $t=2s\pi$, where $s \in \mathbb{N}$ and $2\pi$ is the periodicity of the Hamiltonian.
\xm

When $\epsilon=0$ in Hamiltonian (\ref{eq:H_osc_t_rescale}), the system is integrable and hence gives rise to a regular phase space, where all trajectories reside on equal energy surfaces. According to the KAM (Kolmogorov-Arnol'd-Moser) theorem \cite{book:reichl}, regular features remain in the phase space even when one introduces small integrability breaking perturbation, parametrized by $\epsilon$. The persistence of regular features is shown in the left panels of \fref{phase_space}.
These features gradually get destroyed when $\epsilon$ is increased, and replaced by chaotic trajectories, as we see when comparing the left and the right panels of \fref{phase_space}. A phase space containing chaotic regions with embedded regular islands is known as mixed-phase space.

The key feature of phase-space for the present work are two large period-one islands of regular motion or KAM tori situated symmetrically around $(x,p)=(\pm x_0,0)$ for some $x_0$. These islands are tagged with $\Delta_\pm$ in \frefp{phase_space}{d}. On trajectories within the islands, the mechanical oscillator roughly completes one oscillation when the external modulation completes one, so that on each stroboscopic snapshot the trajectory is found at a similar location. A region of chaotic motion surrounds these islands.  

The KAM theorem \cite{book:reichl} states that a classical trajectory situated in one of the islands $\Delta_\pm$ classically cannot cross into the other island. However, quantum mechanically this statement does not hold, as we shall review nextly.
%
\subsection{Quantum Chaos}
\label{qchoas}

We now move to a quantum description of the dynamics arising from Hamiltonian \bref{eq:H_osc_t_rescale}. Since the Hamiltonian is periodic in time, $\hat{H}(t+T) = \hat{H}(t)$, we can apply Floquet theory \cite{book:reichl}. According to the Floquet theorem, a basis set in the Hilbert space can be found at any given time as
\begin{equation} \label{eq:floq_sol}
    \ket{\chi_n (t)} = \exp(-i E_n t/\heff) \ket{\Phi_n (t)},
\end{equation} 
where $\ket{\Phi_n (t)}$ is a Floquet state, which is periodic with the same period as the Hamiltonian: $\ket{\Phi_n (t+T)} = \ket{\Phi_n (t)}$. $E_n\in \mathbb{R}$ is referred to as quasi-energy. Let us define $\mathcal{\hat{H}}(t) = \hat{H}(t) - i\heff \partial/\partial t$. One finds that
\begin{equation} \label{eq:}
	\mathcal{\hat{H}}(t)\ket{\Phi_n (t)} = E_n \ket{\Phi_n (t)}.
\end{equation}
This shows that $\ket{\Phi_n (t)}$ is an eigenstate of the operator $\mathcal{\hat{H}}(t)$ with eigenvalue $E_n$. Floquet theory allows one to expand the time-evolving state of the system in terms of Floquet states as
\begin{equation}  \label{eq:floq_complete_sol}
    \ket{\Psi (t)} = \sum_n c_n \hspace{1pt} \exp(-i E_n t/\heff) \ket{\Phi_n (t)},
\end{equation}
where the coefficients $c_n$ are set by the initial conditions of the system, akin to the situation for a time-independent Hamiltonian, via $c_n = \braket{\Phi_n(0)}{\Psi(0)}$. 

\begin{figure}[tb]
	\centering
	\includegraphics[width=\linewidth]{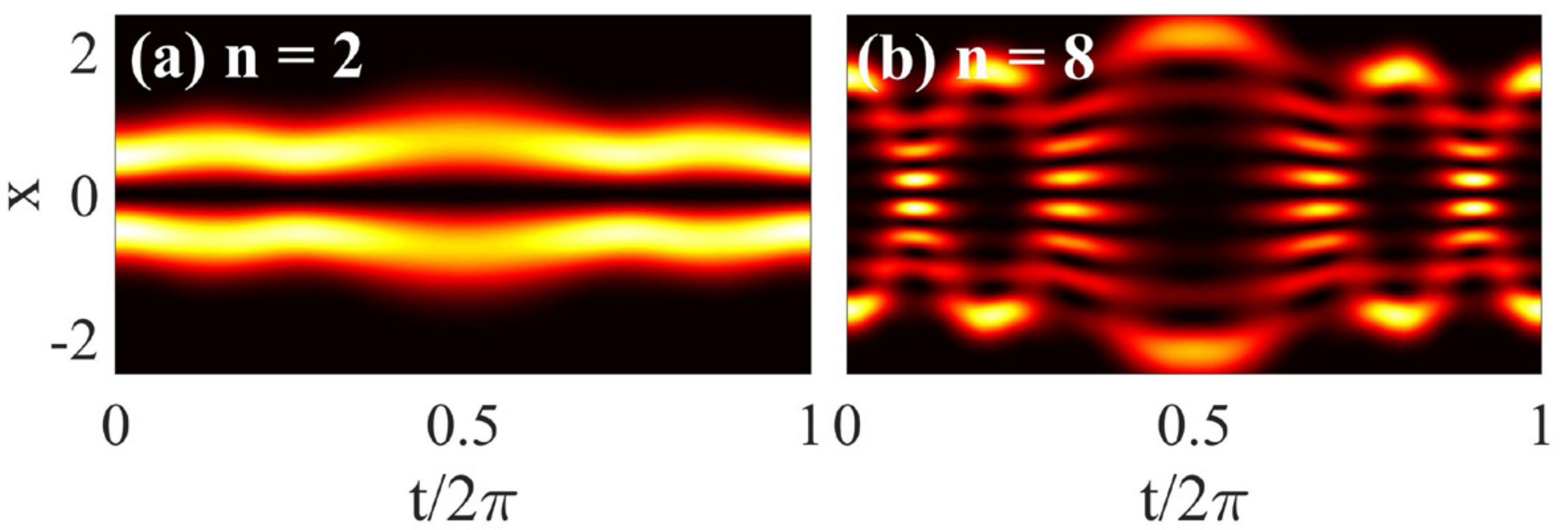}
	\caption{Examples of the evolution of Floquet states (a) $n=2$ and (b) $n=8$ of the Hamiltonian \bref{eq:H_osc_t_rescale} over one period $T=2\pi$ for $\kappa = 1.2$, $\epsilon = 0.9$ and $\heff = 0.5$. The indices $n$ order Floquet states by increasing quasi energy. The colormap represents position-space density $\rho_n(x,t) = \mods{\braket{x}{\Phi_n(t)}}$.
		\label{floquet_states}}
\end{figure}

The periodicity of Floquet states implies that $\ket{\chi_n (t)}$ is reformed after the period $T$, up to some phase. Thus, one obtains $\Phi_n (t=0)$ as eigenstate of the unitary time evolution operator over one period $T$, with complex eigenvalue $\xi_n = \exp (-iE_n T/\heff)$. To construct the evolution operator, we utilise a complete set of symmetric and anti-symmetric position eigenstates as an initial state of the system. The symmetry in the Hamiltonian automatically decouples  the symmetric and anti-symmetric subspaces. We then evolve each eigenstate over one period $T$ according to the time-dependent Schr{\"o}dinger equation that follows from \bref{eq:H_osc_t_rescale}
\begin{equation}  \label{eq:SE_rescale}
    i\heff \frac{\partial {\Psi}}{\partial {t}} =\left[ -\frac{\heff^2}{2} \frac{\partial^2}{\partial {x}^2} + \kappa\frac{{x}^2}{2} + \kappa \big[1+\epsilon \cos ({t}) \big] \frac{{x}^4}{4} \right] {\Psi}.
\end{equation} 
Diagonalization of the resultant time-evolution operator in matrix form, yields the Floquet states $\{ \ket{\Phi_n} \}$ as eigenvectors and the corresponding quasi-energies $\{ E_n \}$ from eigenvalues $\{ \xi_n \}$. \fref{floquet_states} shows the evolution of two selected Floquet states for $\kappa = 1.2$, $\epsilon = 0.9$ and $\heff = 0.5$ over one period of the potential modulation. The required numerical solutions of \bref{eq:SE_rescale} and subsequent ones later in this article are using the high-level code generator XMDS \cite{xmds:docu,xmds:paper}.
%
\begin{figure}[b]
	\centering
	\includegraphics[width=\linewidth]{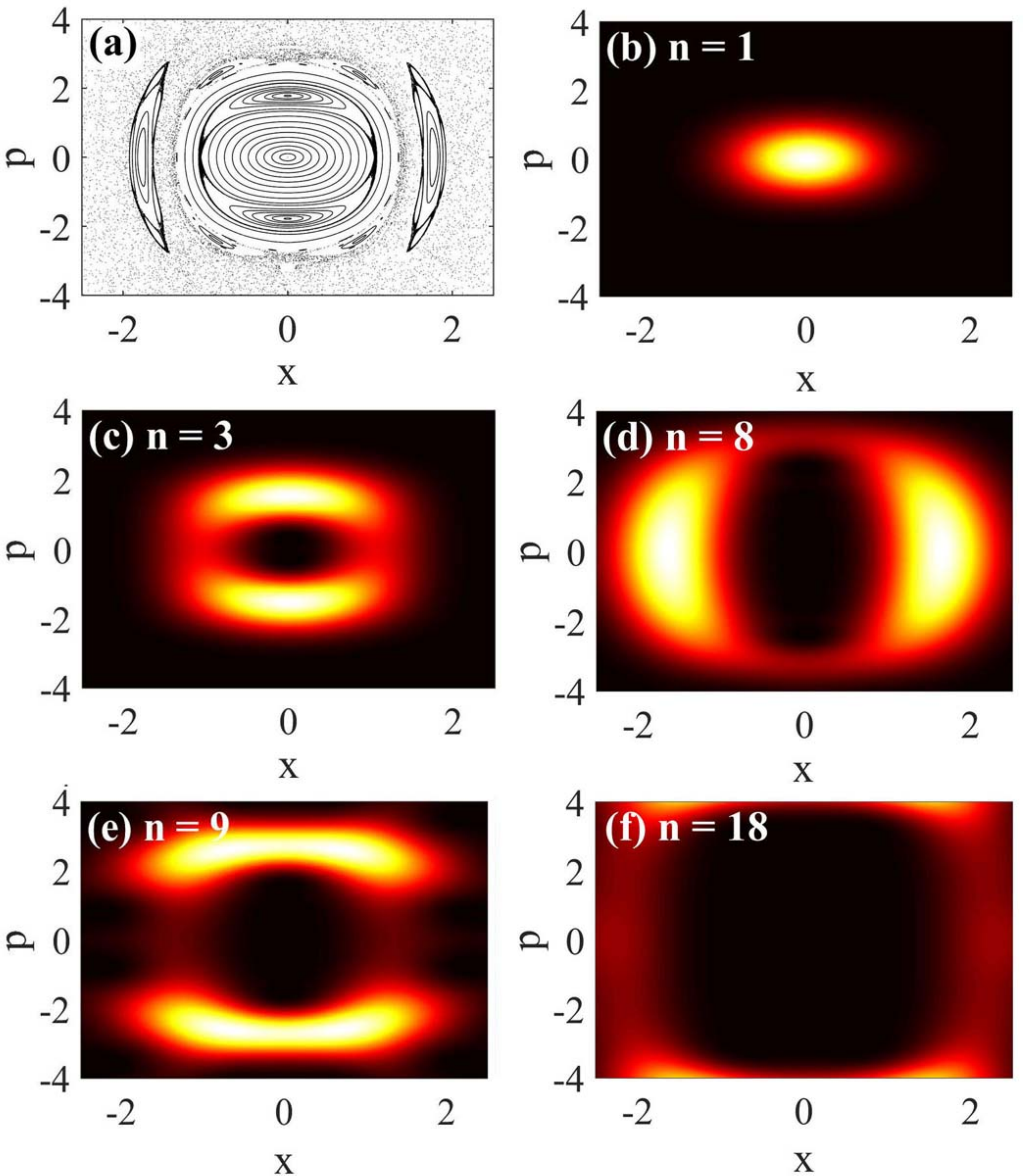}
	\caption{ The support regions of the Husimi distribution of selected Floquet states (b)-(f) for $\kappa = 1.2$, $\epsilon = 0.9$, $\heff = 0.5$ align themselves with features of the classical phase space shown in (a) [identical to \frefp{phase_space}{d}]. Panels (b-f) show $Q(x,p) $ from \bref{eq:husimi} as color shade. These Floquet states are associated with regular (b-d) and chaotic (e,f) regions of phase space. (d) is a tunnelling state defined in the next section.
		\label{husimi_distribution}}
\end{figure}

Now to relate Floquet state to the classical phase spaces in \frefp{phase_space}{d}, we use Husimi (or $Q$) distribution defined as
\begin{equation} \label{eq:husimi}
	Q(x,p) = \frac{1}{2\pi\heff} \mods{\braket{\sub{\alpha}{coh}}{\Phi}},
\end{equation}
where, $\ket{\sub{\alpha}{coh}}$ is a coherent state of the harmonic oscillator centered at position $x$ and momentum $p$. Due to the Heisenberg uncertainty principle, each Floquet state $\ket{\Phi_n}$ must be spread over a finite region of phase space; the extent of which is indicated by the support of $Q$. For $\kappa = 1.2$, $\epsilon = 0.9$ and $\heff = 0.5$, we show the Husimi distribution of selected Floquet states and the classical phase space in \fref{husimi_distribution}, illustrating that Floquet states arrange themselves according to the classical distribution of regular and chaotic regions in phase space. However also note, that most Floquet states have finite overlap with both, regular and chaotic regions. 

\section{Dynamical tunnelling}
\label{dyntun}

As discussed before, a phenomenon that crucially involves several aspects of quantum chaos is dynamical tunnelling \cite{heller:dt}. Recall that motion passing from one of the islands of stability, marked $\Delta_\pm$ in \frefp{phase_space}{d}, to the other is classically forbidden by the KAM theorem. Quantum mechanically, this is no longer true. The period-one regular islands of stability are represented in the Floquet spectrum discussed in \sref{qchoas} by a pair of states covering both islands, with odd / even symmetry under the transformation $x\leftrightarrow-x$ and slightly different quasi energies.
We name those odd ($\ket{\Phi_u}$) and even ($\ket{\Phi_v}$) tunnelling states and identify them as those having maximum overlap with a coherent state centered on the islands. The odd state is shown as example in \frefp{husimi_distribution}{d}. In order to realize a quantum state situated on a single island, we form a linear combination of the tunnelling states
\begin{equation}  \label{eq:tun_initial}
    \ket{\Phi_\pm(0)} = \frac{1}{\sqrt{2}} \big[ \ket{\Phi_u(0)} \pm \ket{\Phi_v(0)} \big],
\end{equation}
where the upper sign locates the state on the right or $\Delta_+$ island. Using the property of Floquet states: $\ket{\Phi_{u,v}(sT)} = \exp (-i E_{u,v} \: sT/\heff) \ket{\Phi_{u,v}(0)}$, the time evolution of the initial state $\ket{\Phi_\pm(0)}$ is
\begin{align}  \label{eq:tun_evol}
\ket{\Phi_\pm(sT)} = \hspace{2pt} &e^{-iE_usT/\heff} \big[ \ket{\Phi_u(0)}  \nonumber\\
&\pm \hspace{2pt} e^{i(E_u- E_v) sT/\heff} \ket{\Phi_v(0)} \big].
\end{align}
The periodic change in the sign of the second term results in transitions between $\ket{\Phi_+}$ and $\ket{\Phi_-}$, which represent dynamical tunnelling. 

\begin{figure}[tb]
	\centering
	\includegraphics[width=0.96\linewidth]{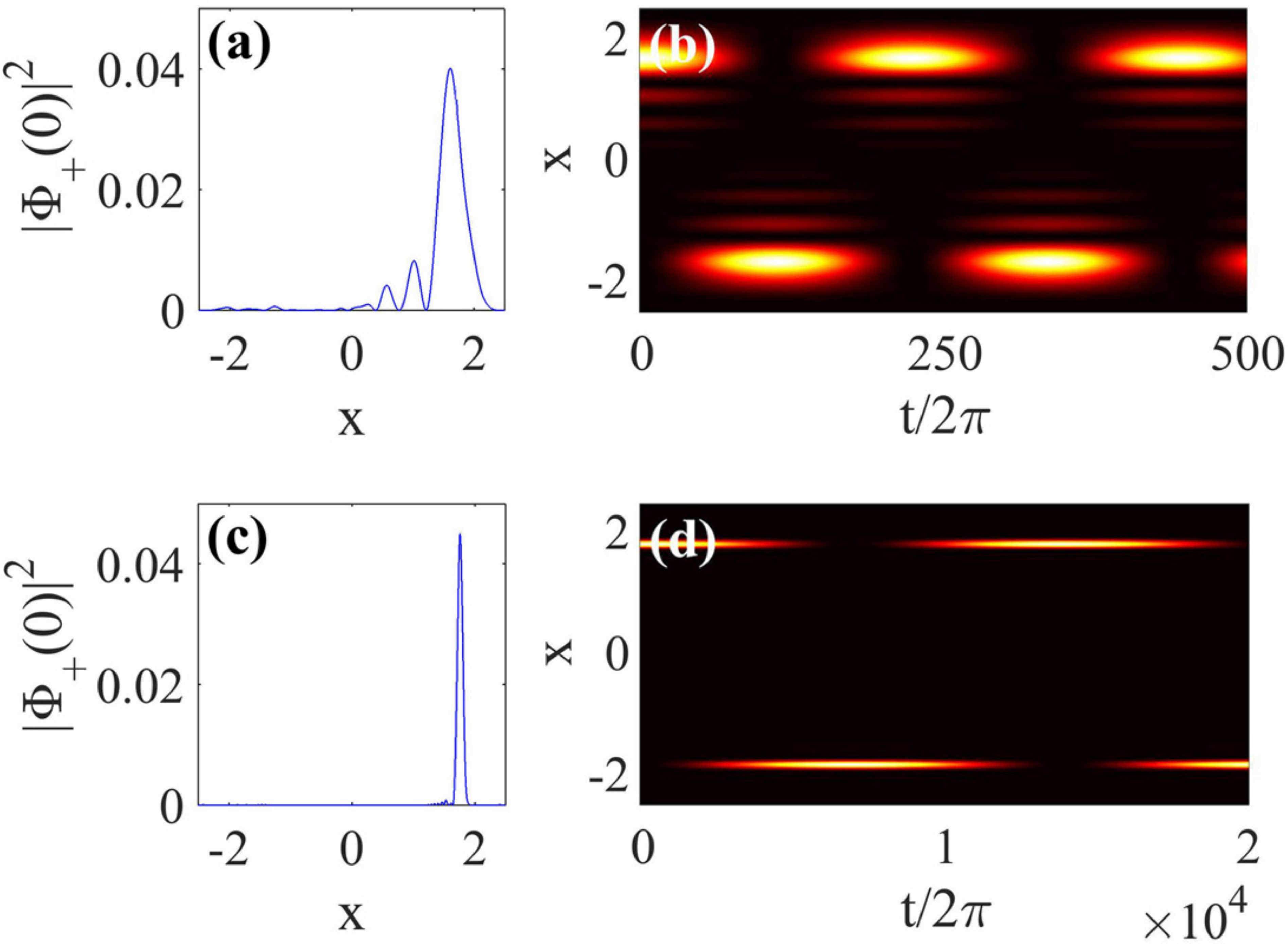}
\caption{ The probability density in the tunnelling state $\ket{\Phi_+}$ at $t=0$ for $\kappa = 1.2$, $\epsilon = 0.9$, and (a) $\heff = 0.5$, (c) $\heff = 0.05$. (b,d) Demonstration of dynamical tunnelling by considering the stroboscopic evolution of the initial states in (a) and (c) in position space. We show the probability density $\rho(x,sT) = \mods{\braket{x}{\Phi_+(sT)}}$ after an integer $s$ of modulation periods $T$, for the parameters as in the corresponding left panel.
\label{tunnelling}}
\end{figure}

We demonstrate this in a direct numerical solution of \bref{eq:SE_rescale}, starting from $\ket{\Phi_+(0)}$, shown in \fref{tunnelling}, for two different values of $\heff$.
The time evolution of the probability density $|\Psi(x,t=2s\pi)|^2$ is again extracted stroboscopically, only after integer modulation periods, at $t=s T$. We see that, unlike the classical case in which the trajectories of the oscillator are confined to their respective islands because of the KAM theorem, the quantum treatment allows population exchange between the symmetry-related islands in a periodic manner. Thus, the Hamiltonian system \bref{eq:H_osc_t_rescale} can show dynamical tunnelling.
%
The period of dynamical tunnelling follows from \bref{eq:tun_evol}, and is controlled by the quasi-energy difference between $\ket{\Phi_u}$ and $\ket{\Phi_v}$  
\begin{equation} \label{eq:tun_period}
	\sub{T}{tun} = \frac{2\pi\heff}{\pbars{E_u- E_v}}.
\end{equation}
$T_{\text{tun}}$ is sensitive to the tunable parameters $\kappa$ and $\epsilon$, see \cite{exp:hensinger}. It increases by orders of magnitude when $\heff$ is reduced as illustrated in \fref{tunnelling} and discussed e.g.~in \cite{narimanov:heffrole,martin:matthew:chip}, reflecting the fact that in the classical limit $\heff\rightarrow 0$ there is no tunnelling. However, for intermediate $\heff$ the tunnelling period can be an interesting probe of the phase-space structure. It has for example been shown that a significant overlap of $\ket{\Phi_{u,v}}$ with the classically chaotic region can again reduce the tunnelling period \cite{utermann:dtperiod} in a phenomenon called chaos-assisted tunnelling. Here the transition of the system from one island of stability to the other can exploit classical transport through the chaotic part of phase space.
%
\section{Creating the Initial Tunnelling State}
\label{inistate}

\begin{figure}[tb]
	\centering
	\includegraphics[width=\linewidth]{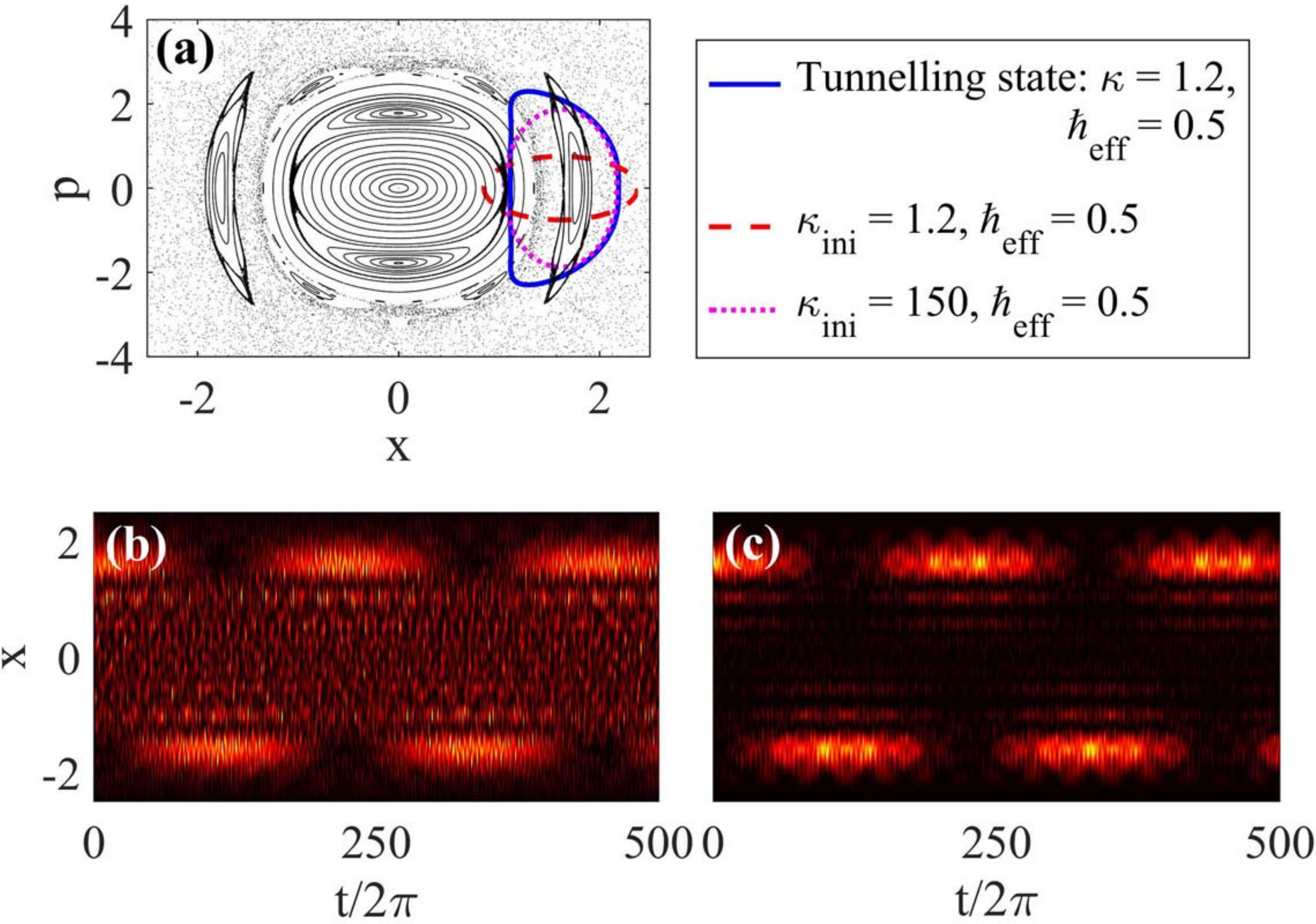}
	\caption{ (a) The Husimi distribution \bref{eq:husimi} of the initial tunnelling state (blue) and two simpler approximations (red, magenta) for $\heff = 0.5$, overlayed as FWHM contour on the classical phase space for the same parameters as in \frefp{phase_space}{d}. The approximations correspond to the ground-state in a modified initial harmonic potential as discussed in the text, with $\sub{\kappa}{ini} = 1.2$ (red dashed) and $\sub{\kappa}{ini} = 150.0$ (magenta dotted). The stroboscopic evolution from the approximate initial tunnelling states is shown in (b) for $\sub{\kappa}{ini} = 1.2$ and (c) for $\sub{\kappa}{ini} = 150.0$.
		\label{ini_tunstate}}
\end{figure}

An exploration of dynamical tunnelling in an optomechanical system will only be possible if the system can at least approximately be brought into the initial tunnelling state in \eref{eq:tun_initial}. As discussed, in \cite{martin:matthew:chip,wuester:selftrapping_long}, this preparation could proceed as follows: We first fit the initial tunnelling state with a coherent state of the harmonic oscillator, having initial position $x_0$, initial momentum $p_0$ and initial width $\sub{\sigma}{ini}$. Since a coherent state is just an oscillator ground-state with an initial kick or position shift, one can create this state by cooling an oscillator to its quantum-mechanical ground-state \cite{Cornell2010:gsmotiondetection, Caltech2011:nanomechosccooling, MIT2011:atomicensamblecooling, EPFL2011:qmgscooling, Boulderl2011:sidebandqmgscooling, delic2020cooling}, and then mechanically offsetting its equilibrium position, or kicking it through the radiation pressure force. 
The resultant state may then be covering an initial island of stability and furnishes an experimentally accessible approximation of the target Floquet state, as shown in \frefp{ini_tunstate}{a}.

We demonstrate in \fref{ini_tunstate} how dynamical tunnelling proceeds from an initial coherent state that is assumed to be the result of the procedure just described. We prepared the approximate initial tunnelling states in \frefp{ini_tunstate}{a} by finding the ground-state in an initially modified potential $\sub{V}{ini}(x) = \sub{\kappa}{ini} x^2/2$ without quartic term or modulation for distinct $\sub{\kappa}{ini}$ and $\heff = 0.5$. Note, that while during ground-state creation the modulation of the quartic term is off-course disabled, one nonetheless can already use the modulation frequency $\Omega$ that will be employed in subsequent time evolution to define time-scale $\tau$ and thence $\heff$.

We see in the stroboscopic evolution shown in \frefp{ini_tunstate}{b} and (c), that follows from these approximate initial island states, that clearly recognizable dynamical tunnelling persists akin to the clean scenario shown in \frefp{tunnelling}{b}. Similar results are shown in \rref{martin:matthew:chip} in the context of dynamical tunnelling in a Bose-Einstein condensate.
\xm

Keeping in mind the scheme above, a central limitation for an experiment will be, whether or not one can provide an initialization potential $\sub{V}{ini}$ sufficiently tight, such that the oscillator will settle into a ground-state with its width matching the width of the Floquet state. One possibility to perform this initial task in the opto-mechanical set-up shown in \frefp{system_sketch}{a} is to set the membrane tilt initially different from the one assumed so far, such that the quadratic coupling between light and membrane is non-zero. Then, one varies the light intensity accordingly to adjust $\sub{\sigma}{ini}$. $\sub{\kappa}{ini}$ in this case can take the form: $\sub{\kappa}{ini} = \kappa(1+\sub{\epsilon}{(2)})$, where the dimensionless parameter $\sub{\epsilon}{(2)}$ involves the quadratic optomechanical interaction and intracavity light field amplitude.
%
\section{Tunable Phase Space Resolution}
\label{ps_resolution}

\begin{figure}[htb]
	\centering
	\includegraphics[width=\linewidth]{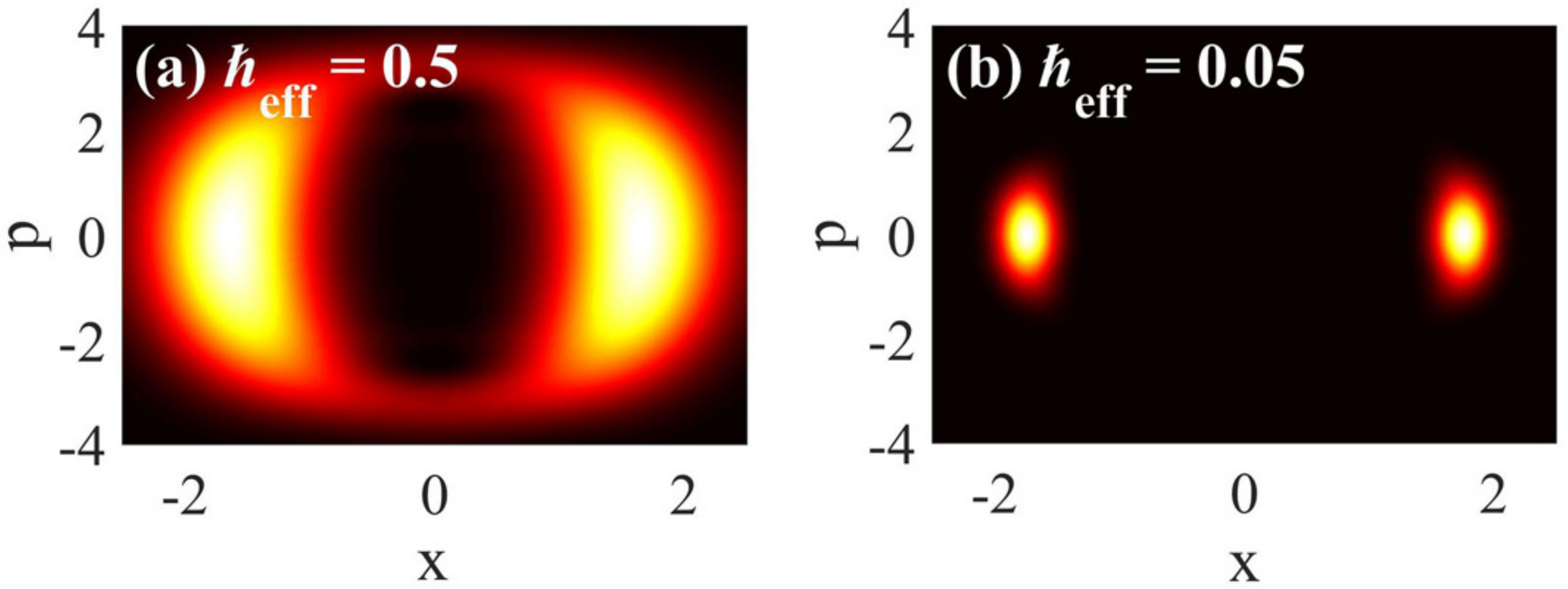}
	\caption{ The Husimi distribution of the odd tunnelling state $\ket{\Phi_u}$ at $t=0$ for $\kappa = 1.2$, $\epsilon = 0.9$, and (a) $\heff = 0.5$, (b) $\heff = 0.05$.
		\label{tuning_heff}}
\end{figure}

A crucial requirement for exploring the quantum-classical boundary in a quantum chaotic system is to be able to continuously vary Hamiltonian parameters such as $\kappa$ and $\epsilon$ to generate different phase-space structures, and then also to be able to reduce $\heff$ and thus turning the system from one with stronger quantum features to one that behaves more classical. \fref{tuning_heff} depicts this process in phase-space, by showing the odd tunnelling state for $\kappa = 1.2$, $\epsilon = 0.9$ and two different $\heff$. It is clear from the figure that a smaller $\heff$ leads to the more localized quantum state, and thus, allows the system to recognize smaller phase space features, as evident from the rescaled Heisenberg uncertainty relation  \eref{eq:commutation_heff}.
\xm
\xm

Typically, practical constraints prohibit a too large variation of all of these parameters in a realistic system. For our specific choice of scaling the variables in \sref{anh_osc_system}, it turns out the clearer obstacle arises when trying to reach a large $\heff$. This is because $\heff$ is proportional to the strength of the quartic contribution to the oscillator potential $g^{(4)}$, which in realistic opto-mechanical systems is typically small. For instance, if we consider the parameters given in \rref{sankey:mim}, with effective oscillator mass $m=50 \hspace{1pt} \text{pg}$, frequency $\omega_m/2\pi=100 \hspace{1pt} \text{kHz}$, quartic coupling to light $\g4/2\pi= 0.4 \hspace{1pt} \text{Hz} \hspace{1pt} \text{nm}^{-4}$, cavity drive laser wavelength $1064 \hspace{1pt} \text{nm}$, laser power $P_0 = 5 \hspace{1pt} \mu\text{W}$, we arrive at an effective Planck's constant of only $\heff \approx 6.7 \times 10^{-15}$ for $\kappa = 1.2$ and $\gamma_c = 10 \hspace{1pt} \omega_m$ using \esref{heff_syspara}{kappa_epsilon_syspara}. However, since $\kappa$ and $\epsilon$ control the size of phase-space features, we need to reach $\heff\approx 1$ in order to genuinely explore the quantum-classical transition and not just classical chaos.

Since the defining parameters of opto-mechanical devices currently available in the field span typically many orders of magnitude \cite{aspelmeyer:review,kippenberg:review}, a comprehensive assessment of how these translate into available $\heff$, $\epsilon$ and $\kappa$ is a major challenge. Instead, we take the parameters quoted above as an initial starting point, and then tweak them in the direction required here. To attain a substantial $\heff$, we then consider $m=1 \hspace{1pt} \text{pg}$, $\omega_m/2\pi=10 \hspace{1pt} \text{kHz}$, $\g4/2\pi= 1 \hspace{1pt} \text{kHz} \hspace{1pt} \text{nm}^{-4}$, laser wavelength $=1064 \hspace{1pt} \text{nm}$, $P_0 = 0.5 \hspace{1pt} \text{mW}$, $\gamma_c = 10 \hspace{1pt} \omega_m$, and $\Omega \approx 0.9 \hspace{1pt} \omega_m$ to keep $\kappa = 1.2$. Altogether, these parameters combine to $\heff \approx 0.042$, which is much larger than the one calculated above. In particular, the quartic coupling strength assumed here has been substantially increased from the one of \rref{sankey:mim}. Since an increase of quartic coupling is widely pursued in the field, also for e.g.~quantum motional state tomography~\cite{weiss2019quantum} or quantum computation~\cite{metzner2020realizing}, we anticipate great progress in this regard.
\s
\begin{figure}[b]
	\centering
	\includegraphics[width=\linewidth]{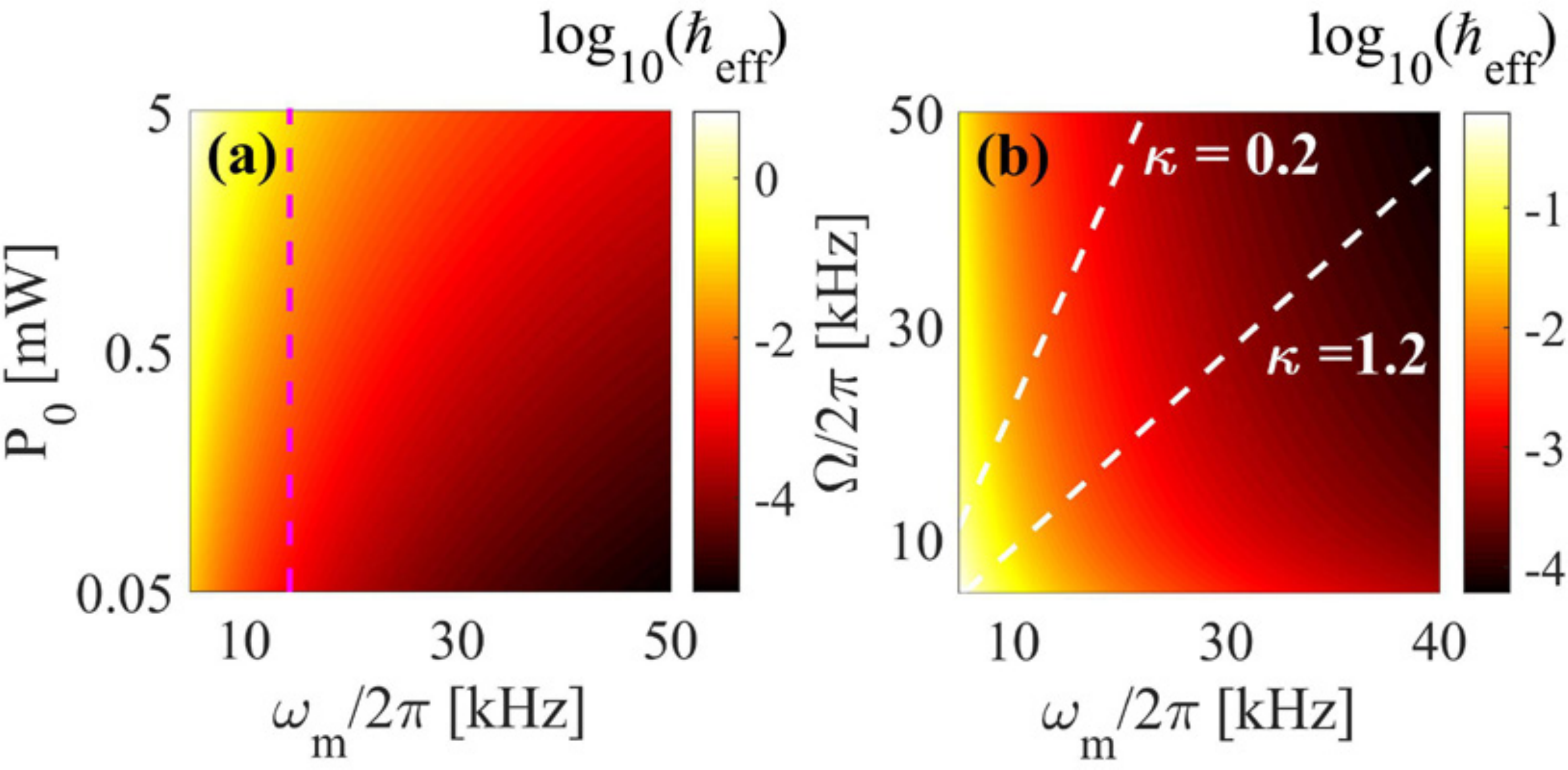}
	\caption{Tunability of the effective Planck's constant $\heff$. (a) As a function of the input laser power $P_0$ and oscillator frequency $\omega_m$, while cavity decay rate $\gamma_c$ and modulation frequency $\Omega$ are kept at a fix ratio with $\omega_m$, see \bref{heff_8a}. Other parameters are held constant at $m=1 \hspace{1pt} \text{pg}$, $\g4/2\pi=1 \hspace{1pt} \text{kHz} \hspace{1pt} \text{nm}^{-4}$, laser wavelength = $1064 \hspace{1pt} \text{nm}$. $\heff \in [0.1,0.001]$ on the magenta line. (b) $\heff$ as a function of the laser field modulation frequency $\Omega$ and oscillator frequency $\omega_m$, while $\gamma_c$ is kept at a fix ratio with $\omega_m$, see \bref{heff_8b}. We use $P_0 = 0.5 \hspace{1pt} \text{mW}$ and other fixed parameters as in (a).
		\label{p0_wm_W}}
\end{figure}

We further show the variation of $\heff$ with two parameters that can be relatively easily adjusted in a single experimental setup in \fref{p0_wm_W}, these parameters being the laser power $P_0$ and its modulation frequency $\Omega$. All other parameters are kept as discussed above. Additional parameter space slices with all relevant equations, including those which are associated with \fref{p0_wm_W}, are discussed in \aref{vary_sys_parameter}. We see that the practically interesting range of $\heff \in (10^{-4}, 1)$ can be covered for example to a large extent on the magenta slice in \frefp{p0_wm_W}{a}. While the estimates above on first sight seem to imply that realizing smaller and smaller $\heff$ is easy in the discussed setup, there are constraints in that direction as well. For smaller $\heff$, the initialization of the oscillator in the tunnelling Floquet state, which has smaller and smaller real space width as $\heff$ is reduced, would become problematic. Apart from this, smaller $\heff$ would give rise to a larger tunnelling period $\sub{T}{tun}$, as seen in \fref{tunnelling}, which eventually becomes challenging to observe.

Since we have neglected decay of the mechanical oscillator, a final limitation will be to have a sufficiently high Q-factor to also cover possibly lengthy dynamical tunneling periods such as in \frefp{tunnelling}{b}.
%

\section{Conclusions and Outlook}
\label{conc_out}
%
We have explored the utility of a membrane in an optical cavity, which furnishes an anharmonic driven quantum oscillator, as a platform for the study of quantum chaos by passing through the quantum to classical transition. Specifically, we focussed on the phenomenon of dynamical tunnelling. A necessary requirement for the realization of chaos in a quantum system with only one spatial degree of freedom is a time-dependent Hamiltonian and an-harmonic potential. In our proposal, both are realized through modulating the field in an optical cavity, in a special setting with significant quartic opto-mechanical coupling proposed in \rref{sankey:mim}. This allows harnessing the advanced control and interrogation tools of quantum opto-mechanics for the study of dynamical tunnelling.

We have shown that the classical phase space for the non-linear oscillator describing the membrane can undergo significant qualitative changes when the 
parameters of its Hamiltonian are varied within experimentally accessible ranges. Moving to quantum mechanics, we then found exemplary Floquet states of this system and simulated dynamical tunnelling. As a central result, we give an overview of accessible variations of the effective Planck's constant $\heff$ as a function of the design parameters of an 
opto-mechanical experiment, and thus demonstrate that the platform is a promising candidate to investigate the quantum to classical transition in a quantum-chaotic system.
Owing to the diversity of quantum opto-mechanical device architectures \cite{kippenberg:review,aspelmeyer:review}, the platform discussed here can complement alternative ones such as Bose-Einstein condensates, which are challenged by practical difficulties in realizing a clean one-dimensional setting \cite{martin:matthew:chip} but in turn offer interesting prospects when dealing with many-body physics \cite{wuester:selftrapping,heimsoth:orbitaljosephson,heimsoth:effectivejosephson,wuester:selftrapping_long,cree:trapping_hamil}.

While the creation of a driven an-harmonic potential for the membrane in the cavity discussed here could be realized with a classical light field in the cavity, which trivially follows the drive laser modulation, non-classical states of light in an optical cavity, coupled to a quantum oscillator, are at the heart of opto-mechanics. An interesting extension of the present work would thus be to consider the dynamics of quantum fluctuations around the mean, and explore their possible coupling to dynamical tunnelling of the intra-cavity non-linear oscillator. We sketch the initial step in this direction in \aref{cavity_quantum_app}, but defer a detailed treatment to future work.
%
\acknowledgments
We are grateful for financial support from the Max-Planck society under the MPG-IISER partner group program. 

\appendix

\section{Cavity field dynamics} \label{cavity_app}

The Hamiltonian \bref{eq:H_system} in principle describes a highly non-trivial system, where the quantum dynamics of a mechanical object couples to that of a light field. This coupling is, of course, at the heart of opto-mechanics. For our present purposes, we only require one of the simplest scenarios, where the two-way coupling usually sought is in fact negligible, and we merely use the cavity as a tool for one-way manipulations of the oscillator. This setting is discussed in \sref{cavity_class_app}. In \sref{cavity_quantum_app} we highlight how one could go beyond this regime in order to furnish interesting two-way coupling between a system exhibiting dynamical tunnelling and a quantum light field, as alluded to in the conclusion.

\subsection{Classical driven field} \label{cavity_class_app}
%
The evolution equation for the cavity field follows from the total Hamiltonian \bref{eq:H_int_system_rotating} as Heisenberg equation for the photon operator $\a(t)$. If we assume the field to be dominated by its mean, we can replace $\a(t) \rightarrow \alpha(t)$, where $\alpha (t) = \expec{\a(t)}$ is just a complex number, and obtain
\begin{equation}  \label{eq:eom_cavityfield}
    \dot{\alpha}(t) = -i ( \sub{\delta}{c} + \g4 x^4 ) \alpha - \frac{\gamma_c}{2} \alpha + \zeta(t).
\end{equation}
Here we have considered decay of the cavity field at rate $\gamma_c$ and neglected other noise sources.

Focussing on the scenario of a bad-cavity, where $\gamma_c$ exceeds all other relevant energy scales, i.e., $\gamma_c >  \pkeys{\omega_m, \g4 \expec{\hat{x}}^4}$, we can neglect all terms not involving either the external laser drive $\zeta(t)$ or the decay-rate and find the formal solution
\begin{equation}
	\alpha(t) = \alpha(0) e^{-\gamma_c t/2} + \int_{0}^{t} ds \hspace{2pt} \zeta(s) e^{-\gamma_c (t-s)/2}.
\end{equation}
Using $\delta(t) = \mbox{lim}_{\epsilon \rightarrow 0} e^{-t/2\epsilon}/2\epsilon$; $t \geq 0$, we reach 
\begin{equation}
	\alpha(t) \approx \alpha(0) e^{-\gamma_c t/2} + \frac{2}{\gamma_c}   \int_{0}^{t} ds  \zeta(s) \delta(t-s) \approx \frac{2}{\gamma_c} \zeta(t),
\end{equation}
so that the cavity field simply follows the external driving amplitude.
This is referred to as adiabatic elimination of the cavity modes \cite{buchmann:adiabatic, agarwal:adiabatic, seok:adiabatic}.

When we consider the periodically-driven laser power $\sub{P}{$\ell$}(t) = P_0 + P_A \cos(\Omega t)$, the cavity field takes the form
\begin{equation}
        \label{eq:simple_modalpha}
	\mods{\alpha(t)} = \mods{\alpha_0} + \mods{A} \cos(\Omega t)
\end{equation}
with the help of $\zeta(t) = \sqrt{2 \sub{P}{$\ell$}(t) \gamma_c/\hbar\sub{\omega}{$\ell$}}$, where
\begin{equation}  \label{eq:alpha0_A}
	\mods{\alpha_0} = \frac{8 P_0}{\hbar \sub{\omega}{$\ell$} \gamma_c} \:\:\:\: \text{and} \:\:\:\: \mods{A} = \frac{8 P_A}{\hbar \sub{\omega}{$\ell$} \gamma_c}.
\end{equation}
We can hence also write $\epsilon=P_A/P_0$. In this simplified picture the intra-cavity field is thus just periodically modulated, with frequency $\Omega$ directly controlled by the external drive laser. We shall assume the simple picture \bref{eq:simple_modalpha} throughout this article.

\subsection{Quantum dynamics} \label{cavity_quantum_app}

Consider the cavity field which fluctuates about its mean $\alpha$ at an amplitude $\delta\a$, such that $\delta\a \ll \alpha$. By linearizing the light field as $\a(t) = \alpha(t) + \delta\a(t)$, the coupled equations of motion for the Hamiltonian \bref{eq:H_int_system_rotating} are
\begin{align} 
   \delta\dot{\a}(t) &= -i ( \sub{\delta}{c} + \g4 \hat{x}^4 ) \delta\a - \frac{\gamma_c}{2} \delta\a + \sqrt{\gamma_c} \hspace{1pt} \delta\sub{\a}{$\ell$}(t),   \label{eq:eom_cavityfluc} \\
   \dot{\hat{x}}(t) &= \frac{\hat{p}}{m} \:\:\: \text{and} \label{eq:xdot_fluc}  \\
   \dot{\hat{p}}(t) &= -m\omega_m^2 \hat{x} -4\hbar\g4 [\mods{\alpha(t)} +  (\alpha^*(t)\delta\a + \alpha(t)\delta\ad)] \hat{x}^3,  \label{eq:pdot_fluc}
\end{align}
where $\delta\sub{\a}{$\ell$}(t)$ is the operator describing the quantum fluctuations of the drive laser. The set of equation \bref{eq:eom_cavityfluc}-\bref{eq:pdot_fluc} makes apparent that quantum fluctuations of the light and mechanical motion and hence dynamical tunnelling are coupled, and thus might exhibit interesting interplay.

\section{Effective rescaled Hamiltonian} \label{nondim}
%
The Schr\"odinger equation following from the Hamiltonian \bref{eq:H_osc} assuming the mean light amplitude \bref{eq:modullight} is
\begin{align} \label{eq:SE_H_osc_t}
	i\hbar\frac{\partial \Psi(x,t)}{\partial t} = \bigg[ &-\frac{\hbar^2}{2m}\frac{\partial^2}{\partial x^2} + \frac{1}{2} m\omega_m^2 x^2 + \hbar \g4 \mods{\alpha_0} x^4  \nonumber \\
	&+ \hbar \g4 \mods{A} \cos(\Omega t) x^4 \bigg] \Psi(x,t).
\end{align}
After defining a time scale $\tau=\Omega^{-1}$ and a length scale $\mathcal{L}=(\sigma \sqrt{8\g4\mods{\alpha_0}/\omega_m})^{-1}$ for the problem, we define the corresponding dimensionless time and space coordinates:
\begin{equation} \label{eq:rescaling}
	\dl{x} = \frac{x}{\mathcal{L}}, \mbox{ and } \dl{t} = \frac{t}{\tau}.
\end{equation}
After conversion to these units, \eref{eq:SE_H_osc_t} takes the form:
\begin{align} \label{eq:SE_rescale_v1}
    i\frac{16\sigma^4 \g4 \mods{\alpha_0}}{\Omega} \frac{\partial \Psi}{\partial \dl{t}} = \bigg[ &-\half \bigg( \frac{16\sigma^4 \g4 \mods{\alpha_0}}{\Omega} \bigg)^2 \frac{\partial^2}{\partial \dl{x}^2} + \frac{\omega_{m}^2}{\Omega^2} \frac{\dl{x}^2}{2} \nonumber \\
    &+ \frac{\omega_{m}^2}{\Omega^2} \bigg( 1+ \frac{\mods{A}}{\mods{\alpha_0}} \cos(\dl{t}) \bigg) \frac{\dl{x}^4}{4} \bigg]\Psi. 
\end{align}
We also define a wave function $\tilde{\Psi}$ that is normalized in the new units, $\int_{-\infty}^{\infty} |\tilde{\Psi}(\tilde{x})|^2d\tilde{x} =1$ and then obtain
\begin{equation}  \label{eq:SE_rescale_app}
    i\heff \frac{\partial \dl{\Psi}}{\partial \dl{t}} = \left[ -\frac{\heff^2}{2} \frac{\partial^2}{\partial \tilde{x}^2} + \kappa\frac{\dl{x}^2}{2} + \kappa \big[1+\epsilon \cos (\dl{t}) \big] \frac{\dl{x}^4}{4} \right] \dl{\Psi},
\end{equation} 
with $\kappa$ and $\epsilon$ as given in \eref{kappa_epsilon}, $\heff$ given in \eref{eq:commutation_heff} and the term in square brackets the effective Hamiltonian given in \bref{eq:H_osc_t_rescale}.

Using \eref{eq:alpha0_A} and the oscillator width $\sigma$, the expression of $\heff$, $\kappa$ and $\epsilon$ in terms of the system parameters is
\begin{align}
	&\heff = \frac{32\hbar\g4 P_0}{m^2\omega_m^2 \Omega \sub{\omega}{$\ell$} \gamma_c},        \label{heff_syspara}   \\  
	\kappa &= \frac{\omega_m^2}{\Omega^2} \:\:\:\:\: \text{and} \:\:\:\:\: \epsilon = \frac{P_A}{P_0}.	\label{kappa_epsilon_syspara}
\end{align}
%
%
\section{Phase space tuning} \label{vary_sys_parameter}

\begin{figure}[htb]
	\centering
	\epsfig{file=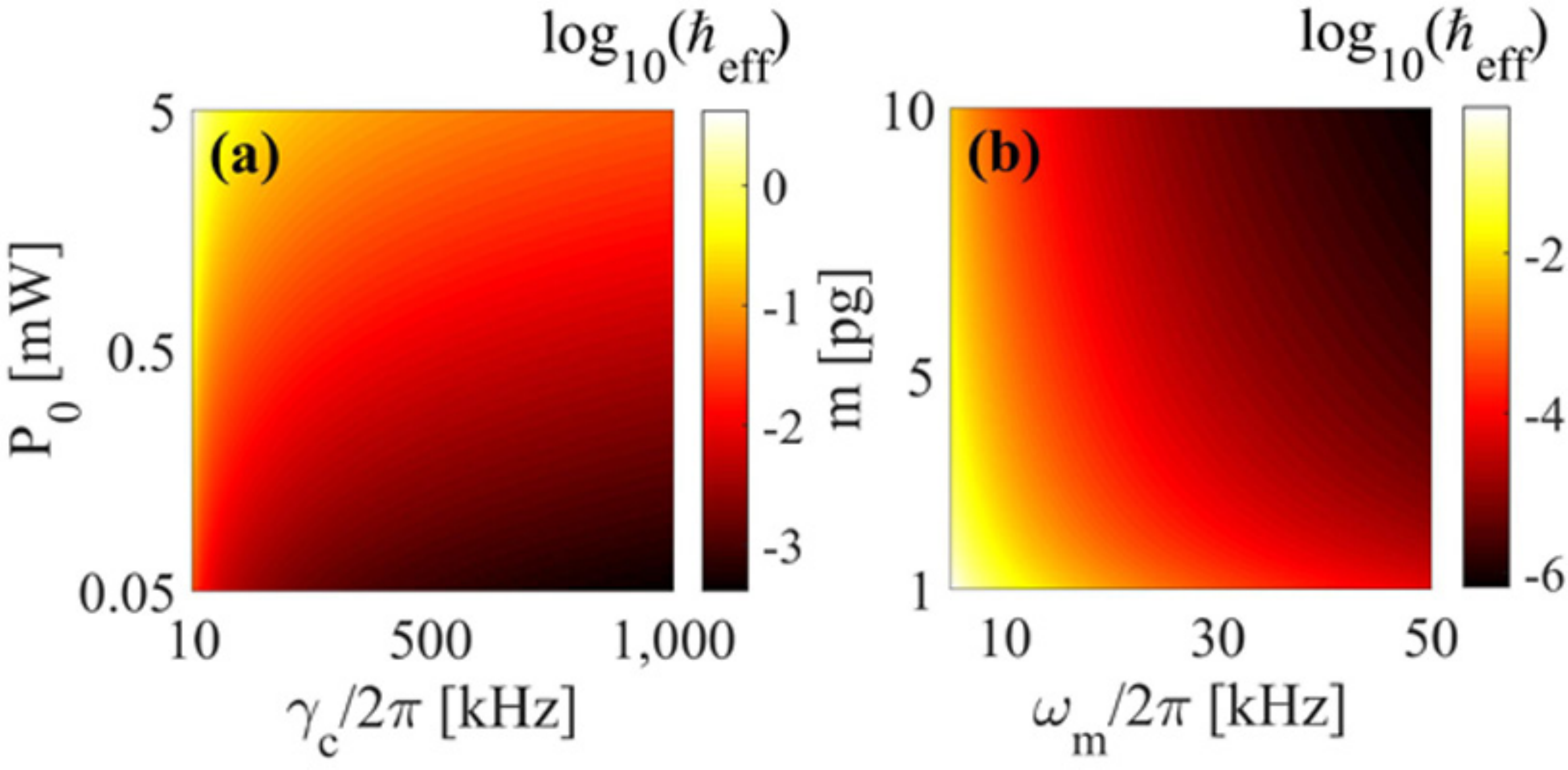,width=\columnwidth} 
	\caption{Tunability of the effective Planck's constant $\heff$ as a function of (a) the input laser power $P_0$ and cavity decay-rate $\gamma_c$, and (b) the oscillator mass $m$ and frequency $\omega_m$. Fixed parameters in (a): $m=1 \hspace{1pt} \text{pg}$, $\omega_m/2\pi=10 \hspace{1pt} \text{kHz}$, $\g4/2\pi=1 \hspace{1pt} \text{kHz} \hspace{1pt} \text{nm}^{-4}$, laser wavelength = $1064 \hspace{1pt} \text{nm}$, see \bref{heff_9a}. Fixed parameters in (b): $P_0 = 0.5 \hspace{1pt} \text{mW}$, $\g4/2\pi=1 \hspace{1pt} \text{kHz} \hspace{1pt} \text{nm}^{-4}$, laser wavelength = $1064 \hspace{1pt} \text{nm}$. $\gamma_c$ and $\Omega$ are varied in sync with $\omega_m$ in (b), see \bref{heff_8a}.
		\label{p0_gc_m_wm}}
\end{figure}

We have shown in \sref{ps_resolution} over which range the effective Planck's constant $\heff$ can be tuned through changing some variable parameters of the setup. 
Here, in \frefp{p0_gc_m_wm}{a} and (b) we additionally illustrate the variation of $\heff$ with the laser parameters $\pkeys{\gamma_c,P_0}$ and oscillator parameters $\pkeys{\omega_m,m}$, respectively.

For \frefp{p0_wm_W}{b}, we begin with \bref{heff_syspara} and then insert our choice $\gamma_c = 10 \hspace{1pt} \omega_m$ from \sref{ps_resolution}, to reach 
\begin{equation} \label{heff_8b}
	\heff = \left( \frac{\hbar\g4 P_0}{5\pi^4 m^2\sub{\omega}{$\ell$}} \right) \frac{1}{(\Omega/2\pi)(\omega_m/2\pi)^3},
\end{equation}
where, the parameters $\Omega$ and $\omega_m$ are treated as variables and the remaining ones as constant. When we insert $\Omega$ from \bref{kappa_epsilon_syspara} into \bref{heff_syspara}, the expression used in \frefp{p0_gc_m_wm}{a} is 
\begin{equation} \label{heff_9a}
	\heff = \left( \frac{16\hbar\g4}{\pi m^2\omega_m^2\sub{\omega}{$\ell$}\Omega} \right) \frac{P_0}{(\gamma_c/2\pi)}. 
\end{equation}

For the figue, $P_0$ and $\gamma_c$ are varied and other parameters held constant. For \frefp{p0_wm_W}{a}, we again fix $\gamma_c = 10 \hspace{1pt} \omega_m$ in \eref{heff_9a} to obtain
\begin{equation} \label{heff_8a}
    \heff = \left( \frac{\hbar\g4\sqrt{\kappa}}{5\pi^4 m^2\sub{\omega}{$\ell$}} \right) \frac{P_0}{(\omega_m/2\pi)^4}, 
\end{equation}
which is also used in  \frefp{p0_gc_m_wm}{b}.

We see from \fref{p0_wm_W} and \ref{p0_gc_m_wm},  that besides the easily tunable parameters like the laser power $P_0$, its modulation frequency $\Omega$ and cavity decay-rate $\gamma_c$, also the less flexible parameters of the mechanical oscillator play a crucial role in deciding the accessible range of $\heff$.

\FloatBarrier
\bibliography{optomech_DT}

\end{document}